\documentclass[12pt]{article}
\usepackage{amsfonts,amsmath}
\usepackage{graphicx,psfrag,epsf}
\usepackage{enumerate}
\usepackage{natbib}
\usepackage{url} % not crucial - just used below for the URL 
\usepackage[dvipsnames]{xcolor}
\usepackage{graphicx}
\usepackage{multirow}
\usepackage{algorithm}
\usepackage{algorithmic}

\newcommand{\blind}{1}

\newcommand{\bX}{\boldsymbol{X}}
\newcommand{\bx}{\boldsymbol{x}}
\newcommand{\bY}{\boldsymbol{Y}}
\newcommand{\mX}{\mathcal{X}}
\newcommand{\bP}{\boldsymbol{P}}
\newcommand{\mbP}{\mathbb{P}}
\newcommand{\mS}{\mathcal{S}}
\newcommand{\btheta}{\boldsymbol{\theta}}
\newcommand{\rd}{\rm{d}}

\begin{document}

\def\spacingset#1{\renewcommand{\baselinestretch}%
{#1}\small\normalsize} \spacingset{1}

%%%%%%%%%%%%%%%%%%%%%%%%%%%%%%%%%%%%%%%%%%%%%%%%%%%%%%%%%%%%%%%%%%%%%%%%%%%%%%

\if1\blind
{
  \title{\bf Efficient Bayesian Model Selection for Coupled Hidden Markov Models with Application to Infectious Diseases}
  \author{Jake Carson\thanks{
  Contact: Jake Carson (Jake.Carson@warwick.ac.uk) Department of Statistics, University of Warwick, Coventry, CV4 7AL, UK. JC and SEFS gratefully acknowledge funding from by MRC grant MR/P026400/1; SEFS was supported by EPSRC grant EP/R018561/1. The authors warmly thank Laurence Tiley for the data used in this article.
   } 
   \hspace{.2cm}\\
    Department of Statistics, University of Warwick, Coventry, UK\\
    Trevelyan J McKinley \\
    College of Medicine and Health, University of Exeter, Exeter, UK\\
    Peter Neal\\
    School of Mathematical Sciences, University of Nottingham, \\ Nottingham, UK\\
    and \\
    Simon EF Spencer \\
    Department of Statistics, University of Warwick, Coventry, UK}
  \maketitle
  
  \newpage
  
} \fi

\if0\blind
{
  \bigskip
  \bigskip
  \bigskip
  \begin{center}
    {\LARGE\bf Efficient Bayesian Model Selection for Coupled Hidden Markov Models with Weak Inter-Chain Dependencies}
\end{center}
  \medskip
} \fi

\bigskip
\begin{abstract}
Performing model selection for coupled hidden Markov models (CHMMs) is highly challenging, owing to the large dimension of the hidden state process.  Whilst in principle the hidden state process can be marginalized out via forward filtering, in practice the computational cost of doing so increases exponentially with the number of coupled Markov chains, making this approach infeasible in most applications.  Monte Carlo methods can be utilized, but despite many remarkable developments in model selection methodology, generic approaches continue to be ill-suited for such high-dimensional problems.  Here we develop specialized solutions for CHMMs with weak inter-chain dependencies.  Specifically we construct effective proposal distributions for the hidden state process that  remain  computationally  viable  as  the  number  of  chains  increases,  and that require little user input or tuning.  This methodology is particularly applicable to individual-level infectious disease models characterized as CHMMs, in which each chain represents an individual, and the coupling represents contact between individuals. Since the only significant contacts are between susceptible and infectious individuals,  and since multiple infection pathways are often possible, the resulting CHMMs naturally have low inter-chain dependencies. We demonstrate the utility of our methodology with an application to a study of highly pathogenic avian influenza in chickens.
\end{abstract}

\noindent%
\vfill

\newpage
\spacingset{1.5} % DON'T change the spacing!

\section{Introduction}
\label{sec:intro}

Hidden Markov models (HMMs) are frequently utilized in the analysis of data that arise from some hidden time-varying state process.
In a HMM this hidden state process is modeled as a Markov chain, and each observation is assumed to only relate to the state of the chain at the associated observation time.
Coupled HMMs \citep[CHMMs;][]{Brand1996} build on the HMM framework in order to model multiple interacting processes.
Each interacting process is modeled as a Markov chain, the dynamics of which, at any given time, depend on the states of every chain at that time.
CHMMs tend to exhibit nonlinear dynamics and unpredictable trajectories, and small changes to the model structure can give rise to radically different outcomes.
Where multiple model specifications can be considered valid, there is a frequent need to select between competing models based on the available data.

There has been a lot of interest within the Bayesian statistics community in developing methods for model selection \citep{Green1995, DelMoral2006, Chib1995, Friel2008, Skilling2004}.
The focus of these approaches is on obtaining Bayes factors, which indicate the strength of evidence between pairs of models.
This is commonly achieved by estimating the model evidence of each model, requiring the integration over all model parameters and hidden states, and from which the Bayes factors can be evaluated.
Despite the many advances in Bayesian model selection methodology, a generic solution for high-dimensional models remains elusive.
However, for certain classes of high-dimensional models it may be possible to construct specialized solutions.

CHMMs typically have relatively few model parameters, which then impose a structure on the evolution of the hidden state process through time.
The challenge for undertaking model selection in CHMMs stems from the need to reconstruct this high-dimensional hidden state process.
In this paper we develop methodology to select between, or average over, competing CHMMs for which the hidden state process is governed by weak inter-chain dependencies.
We construct proposal distributions for the hidden state process, which are then used in an importance sampling algorithm to determine the model evidence of each model.
The key features of our proposal distributions are that they remain computationally viable as the number of individuals being modeled increases, and that they are constructed automatically from the model specification and data, requiring little input from the user.
This allows them to be applied to different models and data sets with ease.

CHMMs are particularly well suited for individual-level infectious disease models:
we are rarely able to detect when an individual becomes infected or recovers, and so modeling the infection status of individuals as a hidden state process is necessary;
by coupling chains we can model disease transmission between individuals;
and diagnostic tests can return false positives or false negatives, which can be accounted for in CHMMs.
Furthermore, inter-chain dependencies tend to be weak as the probability of a susceptible individual becoming infected does not depend on the exact states of all other individuals, but rather some lower dimensional summary providing the infection pressure.
The methodology presented here can be a useful tool for public health professionals to compare the appropriateness of competing models in the light of real-world data.
This can offer crucial insight into disease transmission characteristics, the potential number of future infections, and the effectiveness of possible intervention strategies.

The paper structure is as follows. In Section \ref{S:Models} we describe coupled hidden Markov models and discuss instances of weak inter-chain dependence. In Section \ref{S:Methods} we present our methodology for constructing proposal distributions of the hidden state process and discuss implementation considerations. In Section \ref{S:AIV} we introduce our application data set and provide results. We finish with a discussion in Section \ref{S:Discussion}.

\section{Coupled hidden Markov models}
\label{S:Models}

HMMs comprise a hidden state process $\bX_{1:T}$ that indicates the state of the system over some discrete set of times $t= 1,..,T$, and a visible series of observations $\bY_{1:T}$ that provide imperfect information about the hidden state process.
Elements of the hidden state process are assumed to have a discrete finite state space, such that $\bX_t \in \mX$, where $\mX = \lbrace 1,...,N_X \rbrace$.
The hidden state process is initialized at time $t=1$ with the set of probabilities $\bP_i = \mbP (\bX_{1} = i)$,
and trajectories of the hidden state process are governed by a Markov transition kernel $\bP_{ij} = \mbP (\bX_{t} = j \mid \bX_{t-1} = i)$.
Both $\bP_i$ and $\bP_{ij}$ may additionally depend on model parameters.
The observations may have a discrete or continuous state space, and depend only on the hidden state process at time $t$, i.e.~$p(\bY_t \mid \bX_{1:T}) = p(\bY_t \mid \bX_{t})$. 

CHMMs are a collection of HMMs in which the hidden state processes are linked \citep{Brand1996}. 
We denote $\bX_{t}^{1:K}$ as the states of $K$ chains at time $t$.
The state of chain $k$ at time $t$ depends on the states of all chains at time $t-1$, so that the
trajectories of the hidden state process are governed by Markov transition kernels ${\bP^k_{ij} = \mbP (\bX^k_{t} = j \mid \bX^k_{t-1} = i, \bX^{-k}_{t-1})}$.
An observation for chain $k$ at time $t$ is conditionally independent of the states of other chains given the hidden state $\bX^k_{t}$, i.e. $p(\bY^k_t \mid \bX^{1:K}_{t}) = p(\bY^k_t \mid \bX^k_{t})$.

\begin{figure}[t]
\centering
\includegraphics[width=0.45\textwidth]{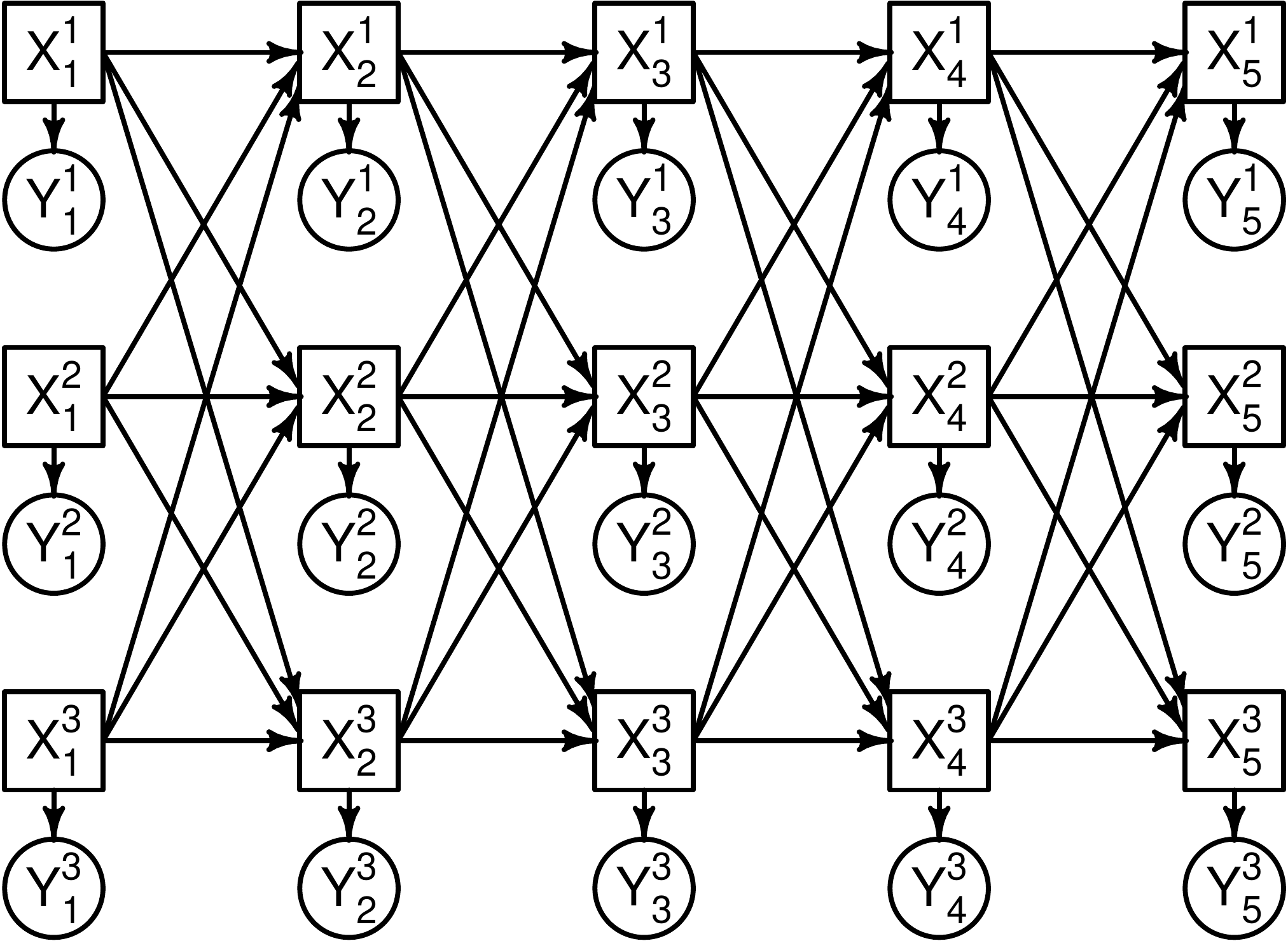} \hspace{30pt}
\includegraphics[width=0.45\textwidth]{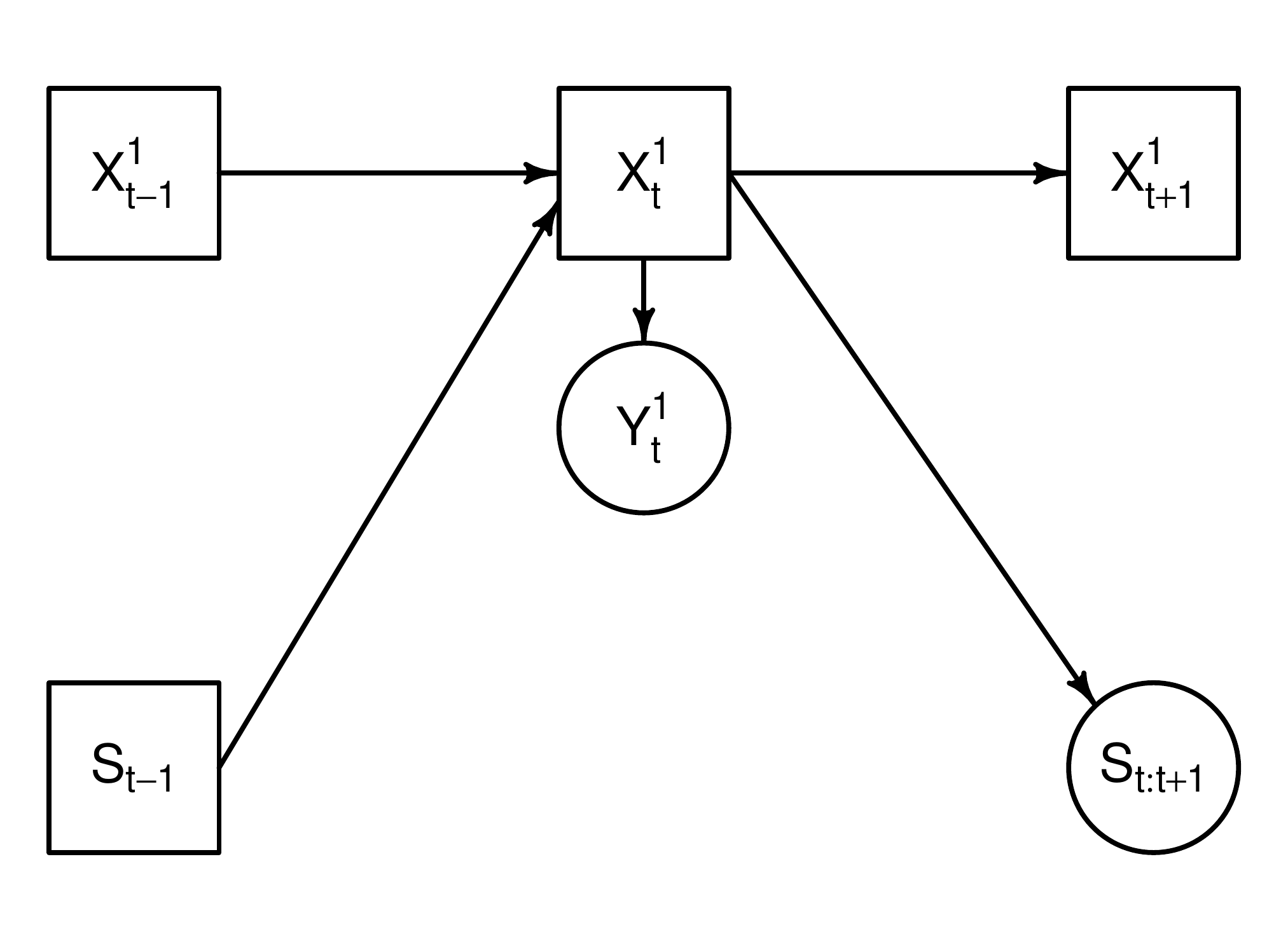}
\caption{Left: An example of a coupled hidden Markov model with three chains. $\bX^k_t$ represent values of the hidden state process, and $\bY^k_t$ represent observations. Superscripts indicate the chain number, and subscripts indicate the time index. Arrows show dependencies between the different variables. Right: An example of a CHMM with weak inter-chain dependencies. With all other values fixed, the value of $\bX^1_t$ depends only on $\bX^1_{t-1}$,  $\bX^1_{t+1}$, $\bY^1_t$, and summary statistics for $\bX^{2:K}_{t-1}$ ($\boldsymbol{S}_{t-1}$) and the transition probabilities of the other chains between times $t$ and $t+1$ ($\boldsymbol{S}_{t:t+1}$). Such CHMMs are commonly used as individual-level infectious disease models.}
\label{F:CHMMCombined}
\end{figure}

In many applications, the Markov transition kernels depend only on a set of summary statistics for the states of the other chains, i.e., ${\bP^k_{ij} = \mbP (\bX^k_{t} = j \mid \bX^k_{t-1} = i, \mS(\bX^{-k}_{t-1}))}$, as shown in Figure~\ref{F:CHMMCombined}.
This is frequently the case when CHMMs are used to model infectious diseases where individuals have been observed longitudinally.
Here, each chain $\bX^k_{1:T}$ represents the hidden state process of individual $k$, and the state space represents the possible set of infection states through which individuals can traverse.
Commonly used infection states include susceptible, infectious, and removed, but others may be used depending on the etiology of the disease.
The more contacts a susceptible individual has with infected individuals, the greater the probability this individual has of becoming infected.
This transmission process between individuals is represented by the coupling of the chains, but is highly abstracted.
An extreme example is when homogeneous population mixing is assumed, in which case the probability of a susceptible individual becoming infected does not depend on which individuals are infectious, only the total number of infectious individuals.
CHMMs with such dependency structures typically exhibit weak inter-chain dependencies, which can be exploited in order to undertake inference using methods that would typically perform poorly in high dimensional problems.

\section{Methods}
\label{S:Methods}

When undertaking Bayesian inference the usual goal is to evaluate the posterior distribution of all unknown parameters and hidden states conditioned on the available data. 
The posterior distribution is given by Bayes' Theorem:
\begin{equation}
\label{E:Post2}
p(\btheta, \bX_{1:T}^{1:K} \mid \bY_{1:T}^{1:K}) \propto p(\bY_{1:T}^{1:K} \mid \bX_{1:T}^{1:K}, \btheta) p (\bX_{1:T}^{1:K} \mid \btheta) p (\btheta),
\end{equation}
\noindent where $p (\btheta)$ is a user defined prior distribution, $p (\bX_{1:T}^{1:K} \mid \btheta)$ is the product of transition probabilities of the hidden state process, and $p(\bY_{1:T}^{1:K} \mid \bX_{1:T}^{1:K}, \btheta)$ is the density induced by the observation model.
In non-trivial situations the posterior distribution is intractable, and so is often numerically approximated using Monte Carlo methods such as MCMC.

When undertaking model selection we are also interested in evaluating the normalizing constant to Equation \eqref{E:Post2}, termed the model evidence.
To evaluate the model evidence we are required to integrate over all model parameters and hidden states:
\begin{equation*}
p(\bY_{1:T}^{1:K}) = \int \int p(\bY_{1:T}^{1:K} \mid \bX_{1:T}^{1:K}, \btheta) p (\bX_{1:T}^{1:K} \mid \btheta) p (\btheta) \; \rd \bX_{1:T}^{1:K} \; \rd \btheta.
\end{equation*}
Taking the ratio of the model evidence terms from two models provides the Bayes factor, which indicates the strength of evidence between the models \citep{Kass1995}.

MCMC does not estimate the model evidence, but many other Monte Carlo approaches have been developed with this purpose in mind.
Some commonly used approaches include importance sampling \citep{Gelfand1994}, SMC \citep{DelMoral2006}, Chib's method \citep{Chib1995, Chib2001, Chen2005},  power posteriors \citep{Friel2008}, and nested sampling \citep{Skilling2004}. 
Such Monte Carlo algorithms rarely perform well when applied to problems with a large number of hidden states.
There are exceptions when the full conditional distribution $p(\bX_{1:T}^{1:K} \mid \bY_{1:T}^{1:K}, \btheta)$ is tractable, or if the (marginal) likelihood $p(\bY_{1:T}^{1:K} \mid \btheta)$ is otherwise available.
For instance, in \citet{Touloupou2018} the full conditional distribution of the hidden state process is incorporated into the proposal distribution in an importance sampling algorithm, and demonstrates that importance sampling can be viable even with a sizable number of hidden states.

When the full conditional distribution $p(\bX_{1:T}^{1:K} \mid \bY_{1:T}^{1:K}, \btheta)$ is intractable, there are no Monte Carlo algorithms that can reliably estimate the model evidence for general CHMMs.
However, for CHMMs with weak inter-chain dependency, we can design effective Monte Carlo solutions to this problem.
The aim of this paper is to provide simple and effective approaches for estimating the model evidence via importance sampling.

\subsection{Importance sampling with hidden states}

For importance sampling we require two proposal distributions, $q(\btheta)$ and $q(\bX^{1:K}_{1:T} \mid \btheta)$, satisfying $q(\btheta) q(\bX^{1:K}_{1:T} \mid \btheta) >0$ whenever $p(\btheta, \bX^{1:K}_{1:T} \mid \bY^{1:K}_{1:T})>0$. Then, for $i=1,...,N$, $\btheta^{(i)}$ and $\bX^{1:K;(i)}_{1:T}$ are sampled from the proposal distributions and assigned importance weights
    \begin{equation*}
    \label{E:IW}
        w^{(i)} = \frac{p\left(\boldsymbol{Y}^{1:K}_{1:T} \mid \boldsymbol{X}^{1:K;(i)}_{1:T}, \boldsymbol{\theta}^{(i)} \right) p\left( \boldsymbol{X}^{1:K;(i)}_{1:T} \mid \boldsymbol{\theta}^{(i)} \right) p\left( \boldsymbol{\theta}^{(i)} \right)}{  q\left( \boldsymbol{X}^{1:K;(i)}_{1:T} \mid \boldsymbol{\theta}^{(i)} \right)  q\left( \boldsymbol{\theta}^{(i)} \right) }. 
    \end{equation*}
\noindent Averaging the importance weights provides an unbiased estimate of the model evidence.

The efficiency of importance sampling is dramatically affected by the choice of proposal distributions, with the optimal choices being $p(\btheta \mid \bY^{1:K}_{1:T})$ and $p(\bX^{1:K}_{1:T} \mid \bY^{1:K}_{1:T}, \btheta)$.
In this case the importance weights equal the model evidence, since we are sampling from the target distribution.
In most situations these distributions are not available, and should instead be approximated.
We discuss suitable approximations for each proposal distribution below.

\subsubsection*{Choice of $q(\btheta)$}

An efficient proposal distribution $q(\btheta)$ will be a good approximation of $p(\btheta \mid \bY^{1:K}_{1:T})$.
Given the success of MCMC algorithms at generating samples from the posterior distribution even when we have hidden states, one possibility is to design proposal distributions based on the output of an MCMC algorithm targeting $p(\btheta \mid \bY^{1:K}_{1:T})$.
\cite{Clyde2007} suggests using a multivariate t-distribution, where the location and scale parameters can be estimated from the output of an MCMC algorithm.
Alternatively, \cite{Hesterberg1995} suggests the use of defense mixtures, for example we could use $q(\btheta) = \lambda \mathcal{N}(\btheta ; \boldsymbol{\mu},\boldsymbol{\Sigma}) + (1-\lambda) p(\btheta)$,
where $\lambda$ is a mixing proportion and $\mathcal{N}(\btheta ; \boldsymbol{\mu},\boldsymbol{\Sigma})$ is a multivariate Gaussian distribution with mean and covariance being estimated from MCMC output.
When the posterior distribution is unimodal, both selections tend to perform well \citep{Touloupou2018}, although the best choice is problem specific.
If the posterior distribution is multi-modal, or otherwise poorly approximated by these distributions, it is possible to fit a mixture of distributions to design the proposal distribution.

\subsubsection*{Choice of $q(\boldsymbol{X}^{1:K}_{1:T} \mid \boldsymbol{\theta})$}

The optimal proposal for the hidden state process is the full conditional distribution $p(\bX^{1:K}_{1:T} \mid \bY^{1:K}_{1:T}, \btheta)$,
in which case the importance weights simplify to 
    \begin{equation*}
        w^{(i)} = \frac{p\left(\boldsymbol{Y}^{1:K}_{1:T} \mid \boldsymbol{\theta}^{(i)} \right) p\left( \boldsymbol{\theta}^{(i)} \right)}{   q\left( \boldsymbol{\theta}^{(i)} \right) }.
    \end{equation*}
\noindent In other words, the hidden state process is marginalized out and we are effectively performing importance sampling only in the parameter space. Since the dimension of the model parameters is usually small compared to that of the hidden state process, the variance of the model evidence estimate will be much lower.

In principle, if the hidden state process takes discrete values then the full conditional distribution can be evaluated and sampled from by using the forward-filtering backward-sampling algorithm (FFBS).
FFBS uses a forward recursion to evaluate the sequence of filtering distributions $p(\boldsymbol{X}^{1:K}_t \mid \boldsymbol{Y}^{1:K}_{1:t}, \boldsymbol{\theta})$ for $t=1,...,T$, followed by a backward recursion to sample from the sequence of distributions $p(\boldsymbol{X}^{1:K}_t \mid \boldsymbol{X}^{1:K}_{t+1}, \boldsymbol{Y}^{1:K}_{1:T}, \boldsymbol{\theta})$.
With each $\boldsymbol{X}^{1:K}_t$ taking $N_X$ possible states, the computational cost of FFBS is $\mathcal{O}(N_X^{2}T)$, leading to poor computational scalability in CHMMs as the number of chains increases.
Specifically, if we assume that each chain has a common state space, such that $\bX_t^k \in \mathcal{S}$, where $\mathcal{S} = \left\lbrace 1,...,S \right\rbrace$, then the number of possible states across all chains is $N_X = S^K$, and the computational cost of FFBS becomes $\mathcal{O}(S^{2K}T)$.
Since the computational cost scales exponentially with $K$, FFBS can only be used in CHMMs with very few chains. 

When the computational cost of performing FFBS is prohibitive, alternative proposal distributions must be considered.
Due to the large dimension of the hidden state process, approximating $p(\bX^{1:K}_{1:T} \mid \bY^{1:K}_{1:T}, \btheta)$ is highly challenging.
Here, we consider adapting methods that have been utilized in MCMC to update the hidden state process conditional on a set of parameter values.
Many such methods, for example single-site updates \citep{Dong2012} and block updates \citep{Spencer2015}, are computationally fast, but can cause slow mixing in MCMC.
Recently the Individual FFBS algorithm \citep[IFFBS;][]{Touloupou2020scalable} was developed, which is computationally scalable in the number of chains, and maintains strong mixing in MCMC for CHMMs with weak inter-chain dependencies.

IFFBS updates the hidden state process using Gibbs sampling on each chain. 
That is, sampling from the distributions  ${p\left(\boldsymbol{X}^{k}_{1:T} \mid \boldsymbol{X}^{-k}_{1:T}, \boldsymbol{Y}^{k}_{1:T}, \boldsymbol{\theta}\right).}$
This is achieved by sequentially performing a modified version of FFBS on each chain, conditional on the states of all other chains through time. 
The computational cost of IFFBS is $\mathcal{O}(S^3 K T)$, which is linear in number of chains and a significant improvement over FFBS.
In MCMC, IFFBS can cause poor mixing when chains are highly correlated, but this tends to be uncommon for CHMMs with weak inter-chain dependencies. 
Here we focus on further developing IFFBS for use in importance sampling.

\subsection{Using IFFBS to design importance proposal distributions}

We develop two proposal distributions that utilize IFFBS.
For each proposed parameter $\btheta^{(i)}$, we obtain a set of samples $\widehat{\boldsymbol{X}}_{1:T}^{1:K;(j)}$, $j=1,\dots,N$ distributed as $p(\bX^{1:K}_{1:T} \mid \bY^{1:K}_{1:T}, \btheta)$ using IFFBS.
Our proposal distributions are then constructed based on these \textit{guiding samples}.

\subsubsection*{Direct application of IFFBS}

The simplest approach is to use IFFBS itself as a proposal distribution, initiated using the guiding samples.
We call this direct IFFBS (DIFFBS).
We determine a high posterior probability realization of the hidden state process from the guiding samples, denoted $\widetilde{\bX}^{1:K}_{1:T}$, and then propose from 
\begin{multline}
\label{E:DIFFBSProp}
q(\bX^{1:K}_{1:T} \mid \btheta^{(i)}) = p(\bX^1_{1:T} \mid \widetilde{\bX}^{2:K}_{1:T}, \bY^{1}_{1:T}, \btheta^{(i)}) \prod_{k=2}^{K-1}   p(\bX^k_{1:T} \mid \bX^{1:k-1}_{1:T}, \widetilde{\bX}^{k+1:K}_{1:T}, \bY^{k}_{1:T}, \btheta^{(i)}) \\
 \times p(\bX^K_{1:T} \mid \bX^{1:K-1}_{1:T}, \bY^{K}_{1:T}, \btheta^{(i)}).
\end{multline}
Here, each term in the product is an IFFBS update step for a single chain.
Proposing one realization of the hidden state process in this manner is computationally expensive, as obtaining the guiding samples requires $N$ iterations of IFFBS over every chain.
Instead, we can propose multiple realizations of the hidden state process for each $\btheta^{(i)}$ based on the same guiding samples and then use the following importance weight:
\begin{equation*}
\label{E:IW2}
    w^{(i)} = \frac{p\left( \boldsymbol{\theta}^{(i)} \right)}{ q\left( \boldsymbol{\theta}^{(i)} \right) } \frac{1}{L} \sum_{l=1}^{L} \frac{p\left(\boldsymbol{Y}^{1:K}_{1:T} \mid \boldsymbol{X}^{1:K;(i,l)}_{1:T}, \boldsymbol{\theta}^{(i)} \right) p\left( \boldsymbol{X}^{1:K;(i,l)}_{1:T} \mid \boldsymbol{\theta}^{(i)} \right) }{  q\left( \boldsymbol{X}^{1:K;(i,l)}_{1:T} \mid \boldsymbol{\theta}^{(i)} \right)  }, 
\end{equation*}
This can be performed using the same realization $\widetilde{\bX}^{1:K}_{1:T}$ for each of the $L$ proposal distributions, or alternatively we could use $L$ different realizations from the guiding samples.

Whilst DIFFBS has the benefit of being a simple adaptation of IFFBS, it does not necessarily provide a good approximation of the optimal proposal distribution.
Consider decomposing the optimal proposal distribution by chain,
\begin{equation}
\label{E:OptProp}
p(\bX^{1:K}_{1:T} \mid \bY^{1:K}_{1:T}, \btheta^{(i)}) = p(\bX^1_{1:T} \mid \bY^{1:K}_{1:T}, \btheta^{(i)}) \prod_{k=2}^{K}   p(\bX^k_{1:T} \mid \bX^{1:k-1}_{1:T},\bY^{k:K}_{1:T}, \btheta^{(i)}),
\end{equation}
\noindent and then comparing this with Equation \eqref{E:DIFFBSProp}.
It is clear that we are proposing from a series of conditional distributions in place of marginal distributions, i.e. a comparison between ${p(\bX^k_{1:T} \mid \bX^{1:k-1}_{1:T}, \widetilde{\bX}^{k+1:K}_{1:T}, \btheta^{(i)}, \bY^{k}_{1:T})}$ and ${p(\bX^k_{1:T} \mid \bX^{1:k-1}_{1:T}, \btheta^{(i)},\bY^{k:K}_{1:T})}$ shows that we are conditioning on values for $\bX^{k+1:K}_{1:T}$ instead of marginalizing them out.
This will likely lead to under-dispersed proposal distributions, and high variance model evidence estimates.

\subsubsection*{Numerical marginalization via IFFBS}

Here we construct a second algorithm based on the decomposition of the optimal proposal distribution given in Equation \eqref{E:OptProp}.
The basic principle is to use the guiding samples to obtain Monte Carlo estimates of each marginal distribution using IFFBS.
As such, we refer to this algorithm as marginal IFFBS, or MIFFBS.

For the first term, $p(\boldsymbol{X}^{1}_{1:T} \mid \boldsymbol{Y}^{1:K}_{1:T} , \boldsymbol{\theta} )$, we decompose in a similar manner as FFBS,
\begin{equation}
\label{E:OptDecom}
p( \boldsymbol{X}^{1}_{1:T} \mid \boldsymbol{Y}^{1:K}_{1:T} , \boldsymbol{\theta} ) = p(\bX^{1}_T \mid \bY^{1:K}_{1:T}, \btheta) \prod_{t=1}^{T-1} p(\bX^{1}_t \mid \bX^{1}_{t+1:T}, \bY^{1:K}_{1:T}, \btheta).
\end{equation} 
\noindent Note that we no longer have the Markov property, as we are marginalizing out the other chains.
The first term on the right hand side of Equation \eqref{E:OptDecom} can be expressed as
\begin{equation*}
\label{E:OptDecomP1} p(\bX^{1}_T \mid \bY^{1:K}_{1:T}, \btheta) = \int p(\bX^{1}_T \mid \bX_{1:T}^{2:K}, \bY^{1}_{1:T}, \btheta) p(\bX_{1:T}^{2:K} \mid \bY^{1:K}_{1:T}, \btheta) \rd \bX_{1:T}^{2:K},
\end{equation*} 
for which we can obtain a Monte Carlo approximation by sampling from ${p(\bX_{1:T}^{2:K} \mid \bY^{1:K}_{1:T}, \btheta)}$ and giving each sample a weight proportional to $p(\bX^{1}_T \mid \bX_{1:T}^{2:K}, \bY^1_{1:T}, \btheta)$. 
To obtain such a sample, we can take our guiding samples and discard the values for the first chain, leaving $\widehat{\bX}_{1:T}^{2:K;(j)}$, $j=1,...,N$.
An approximation to $p(\bX^{1}_T \mid \bY^{1:K}_{1:T}, \btheta)$ is then given by
\[ \widehat{p}(\bX^{1}_T \mid \bY^{1:K}_{1:T}, \btheta) = \frac{1}{N}\sum_{j=1}^{N} p(\bX^{1}_T \mid \widehat{\bX}_{1:T}^{2:K;(j)}, \bY^1_{1:T}, \btheta). \]
\noindent Each component of the sum
is obtained from the forward recursion of IFFBS conditioned on $\widehat{\bX}_{1:T}^{2:K;(j)}$.
We then propose $\bx^{1}_{T} \sim \widehat{p}(\bX^{1}_T \mid \bY^{1:K}_{1:T}, \btheta)$ and initialize the backward sampling recursion of MIFFBS for the first chain. 

In the backward recursion we need to approximate the terms in the product of Equation \eqref{E:OptDecom}, which are 
conditioned on parts of the hidden state process that have already been proposed. 
These terms are again marginal distributions that can be expressed as
\[ p(\bX^{1}_t \mid \bx^{1}_{t+1:T}, \bY^{1:K}_{1:T}, \btheta) = \int p(\bX^{1}_t \mid \bx^{1}_{t+1:T}, \bX^{2:K}_{1:T}, \bY^{1}_{1:t}, \btheta) p(\bX^{2:K}_{1:T} \mid \bx^{1}_{t+1:T}, \bY^{1:K}_{1:T}, \btheta) \rd \bX^{2:K}_{1:T} . \]
\noindent In order to construct Monte Carlo approximations as before, we could consider obtaining samples from ${p(\bX^{2:K}_{1:T} \mid \bx^{1}_{t+1:T}, \bY^{1:K}_{1:T}, \btheta)}$ and giving each sample a weight proportional to ${p(\bX^{1}_t \mid \bx^{1}_{t+1:T}, \bX^{2:K}_{1:T}, \bY^{1}_{1:t}, \btheta)}$.
However, this will be computationally expensive to undertake at every time step.
Instead, note that
\[ p(\bX^{2:K}_{1:T} \mid \bx^{1}_{t+1:T}, \bY^{1:K}_{1:T}, \btheta) \propto p(\bx^{1}_{t+1:T} \mid \bX^{2:K}_{1:T}, \bY^{1:K}_{1:T}, \btheta) p(\bX^{2:K}_{1:T} \mid \bY^{1:K}_{1:T}, \btheta), \]
\noindent suggesting that we can make use of the existing guiding samples, which are distributed according to $p(\bX^{2:K}_{1:T} \mid \bY^{1:K}_{1:T}, \btheta)$.
Hence we can use the Monte Carlo approximation 
\[ \widehat{p}(\bX^{1}_t \mid \bx^{1}_{t+1:T}, \bY^{1:K}_{1:T}, \btheta) \propto \sum_{j=1}^{N} p(\bX^{1}_t \mid \bx^{1}_{t+1:T}, \widehat{\bX}_{1:T}^{2:K;(j)}, \bY^{1}_{1:t}, \btheta) p(\bx^{1}_{t+1:T} \mid \widehat{\bX}_{1:T}^{2:K;(j)}, \bY^{1}_{1:T}, \btheta), \]
\noindent in order to propose $\bx^{1}_{t}$. 
Since we are approximating a probability mass function on a finite state-space, we can easily self-normalize.
The term ${p(\bX^{1}_t \mid \bx^{1}_{t+1:T}, \widehat{\bX}_{1:T}^{2:K;(j)}, \bY^{1}_{1:t}, \btheta)}$ is obtained by performing a backward step of IFFBS between $t+1$ and $t$. The term $p(\bx^{1}_{t+1:T} \mid \widehat{\bX}_{1:T}^{2:K;(j)}, \bY^{1}_{1:T}, \btheta)$ can be decomposed as 
\begin{multline*} 
p(\bx^{1}_{t+1:T} \mid \widehat{\bX}_{1:T}^{2:K;(j)}, \bY^{1}_{1:T}, \btheta) = \\ p(\bx^{1}_T \mid \widehat{\bX}_{1:T}^{2:K;(j)}, \bY^{1}_{1:T}, \btheta) \prod_{u=t+1}^{T-1} p(\bx^{1}_{u} \mid \bx^{1}_{u+1:T}, \widehat{\bX}_{1:T}^{2:K;(j)}, \bY^{1}_{1:u}, \btheta),
\end{multline*}
\noindent the components of which have already been calculated as part of the backwards recursion.
Repeating this process backwards through time gives a set of proposed values $\bx^{1}_{1:T}$ approximately distributed according to $p(\boldsymbol{X}^{1}_{1:T} \mid \boldsymbol{Y}^{1:K}_{1:T} , \boldsymbol{\theta} ) $.

Having proposed values for the first chain, we now wish to propose values for the remaining chains sequentially.
That is, we wish to approximate 
\[ p( \boldsymbol{X}^k_{1:T} \mid \boldsymbol{x}^{1:k-1}_{1:T}, \boldsymbol{Y}^{k:K}_{1:T} , \boldsymbol{\theta} ) = p( \boldsymbol{X}^k_{T} \mid \boldsymbol{x}^{1:k-1}_{1:T}, \boldsymbol{Y}^{k:K}_{1:T} , \boldsymbol{\theta} ) \prod_{t=1}^{T} p( \boldsymbol{X}^k_{t} \mid \boldsymbol{X}^k_{t+1:T}, \boldsymbol{x}^{1:k-1}_{1:T}, \boldsymbol{Y}^{k:K}_{1:t} , \boldsymbol{\theta} ). \]
Again,  each of these terms are marginal distributions, and so for $k<K$
\begin{multline*}
p( \boldsymbol{X}^k_{T} \mid \boldsymbol{x}^{1:k-1}_{1:T}, \boldsymbol{Y}^{k:K}_{1:T} , \boldsymbol{\theta} ) = \\ \int p( \boldsymbol{X}^k_{T} \mid \boldsymbol{x}^{1:k-1}_{1:T}, \boldsymbol{X}^{k+1:K}_{1:T}, \boldsymbol{Y}^k_{1:T} , \boldsymbol{\theta} ) p( \boldsymbol{X}^{k+1:K}_{1:T} \mid \boldsymbol{x}^{1:k-1}_{1:T}, \boldsymbol{Y}^{k:K}_{1:T}, \boldsymbol{\theta} ) \rd \bX_{1:T}^{k+1:K}.
\end{multline*}
\noindent Considering a Monte Carlo approximation, one approach would be to obtain samples ${p( \boldsymbol{X}^{k+1:K}_{1:T} \mid \boldsymbol{x}^{1:k-1}_{1:T}, \boldsymbol{Y}^{k:K}_{1:T}, \boldsymbol{\theta} )}$ that are given weights $p( \boldsymbol{X}^k_{T} \mid \boldsymbol{x}^{1:k-1}_{1:T}, \boldsymbol{X}^{k+1:K}_{1:T}, \boldsymbol{Y}^k_{1:T} , \boldsymbol{\theta} ).$
However, at this stage we have a weighted sample $\widehat{\bX}_{1:T}^{k:K}$ approximating $p( \boldsymbol{X}^{k:K}_{1:T} \mid \boldsymbol{x}^{1:k-1}_{1:T}, \boldsymbol{Y}^{k:K}_{1:T}, \boldsymbol{\theta} )$.
It is computationally cheaper to simply discard all samples for chain $k$, and treat the remainder of the weighted sample as being representative of $p( \boldsymbol{X}^{k+1:K}_{1:T} \mid \boldsymbol{x}^{1:k-1}_{1:T}, \boldsymbol{Y}^{k:K}_{1:T}, \boldsymbol{\theta} )$. 
Proposing values for $\bx^k_{1:T}$ then proceeds as in the case for chain 1.
Finally, for $k=K$ we can sample from the full conditional distribution $p( \boldsymbol{X}^K_{1:T} \mid \boldsymbol{x}^{1:K-1}_{1:T}, \boldsymbol{Y}^K_{1:T} , \boldsymbol{\theta} )$ using IFFBS. 

As with many other algorithms that use sequential weighted approximations to characterize distributions of interest, degeneracy issues may occur.
Similar to SMC, this can be tracked by monitoring the effective sample size (ESS) of the guiding samples.
It makes sense to regenerate the guiding samples when the ESS falls below some threshold, say $\frac{N}{2}$, as this indicates that the approximation is being dominated by a small number of samples.
To do so we generate a new set of guiding samples from scratch using IFFBS conditioned on any values of the hidden state process we have proposed thus far.
Since we only need to obtain guiding samples for chains for which we do not have proposed values, this regeneration step becomes computationally cheaper as the algorithm progresses.
The full algorithm is presented in Algorithm 1.

\begin{algorithm}[p]

{ \spacingset{1} 
{\footnotesize

\caption{MIFFBS algorithm.}
\label{A:FFBS}

{

\begin{algorithmic}

\STATE{Obtain sample $\widehat{\bX}_{1:T}^{1:K;(1:N)}$ distributed as $p(\bX_{1:T}^{1:K} \mid \bY_{1:T}^{1:K}, \btheta)$ using IFFBS, and set weights $W^{(n)} = \frac{1}{N}$.}
\FOR{$k=1,...,K$}
\STATE{- If $\mbox{ESS}<\frac{N}{2}$ obtain a new sample $\widehat{\bX}_{1:T}^{k:K;(1:N)}$ distributed as $p(\bX_{1:T}^{k:K} \mid \bx_{1:T}^{1:k-1}, \bY_{1:T}^{k:K}, \btheta)$ using IFFBS, and set weights $W^{(n)} = \frac{1}{N}$.}
\FOR{$t=1,..,T$}
\STATE{- Calculate modified forward filtered probabilities $p(\bX_t^{k} \mid \bx_{1:t+1}^{1:k-1},\widehat{\bX}_{1:t+1}^{k+1:K;(n)}, \bY_{1:t}^{k}, \btheta )$ for $\bX_t^{k} \in \mathcal{S}$ and $n=1,...,N$ (see Algorithm \ref{A:MFFP}).
}
\ENDFOR
\STATE{- Calculate weighted average of the final modified forward filter probabilities for $\bX_T^{k} \in \mathcal{S}$
\[
\widehat{p} (\bX_T^{k} \mid \bx_{1:T}^{1:k-1}, \bY_{1:T}^{k:K}, \btheta) = \sum_{n=1}^{N} W^{(n)} p(\bX_T^{k} \mid \bx_{1:T}^{1:k-1},\widehat{\bX}_{1:T}^{k+1:K;(n)}, \bY_{1:T}^{k}, \btheta ).
\]
}
\STATE{- Sample $\bx_T^{k}$ from $\widehat{p} (\bX_T^{k} \mid \bx_{1:T}^{1:k-1}, \bY_{1:T}^{k:K}, \btheta)$ and  for $n=1,...,N$ set weight \[  
\mbox{ $ W^{(n)} \propto W^{(n)}  p(\bx_T^{k} \mid \bx_{1:T}^{1:k-1},\widehat{\bX}_{1:T}^{k+1:K;(n)}, \bY_{1:T}^{k}, \btheta ). $ } \] }
\FOR{$t=T-1,...,1$}
\STATE{- Calculate the backward smoothing probabilities for $\bX_t^{k} \in \mathcal{S}$ and $n=1,...,N$ 
\begin{multline*}
 p(\bX_t^{k} \mid \bx_{1:T}^{1:k-1}, \bx_{t+1:T}^{k},  \widehat{\bX}_{1:T}^{k+1:K;(n)}, \bY_{1:T}^{k}, \btheta ) = \\
  \frac{p(\bX_t^{k} \mid \bx_{1:t+1}^{1:k-1},\widehat{\bX}_{1:t+1}^{k+1:K;(n)}, \bY_{1:t}^{k}, \btheta ) p(\bx_{t+1}^{k} \mid \bx_{t}^{1:k-1},\bX_{t}^{k},\widehat{\bX}_{t}^{k+1:K;(n)},\btheta ) }{p(\bx_{t+1}^{k} \mid \bx_{1:t+1}^{1:k-1},\widehat{\bX}_{1:t+1}^{k+1:K;(n)}, \bY_{1:t}^{k}, \btheta )} 
\end{multline*}
}
\STATE{- Calculate weighted average of the backward smoothing probabilities $\bX_t^{k} \in \mathcal{S}$
\[ 
\widehat{p} (\bX_t^{k} \mid \bx_{1:T}^{1:k-1}, \bx_{t+1:T}^{k}, \bY_{1:T}^{k:K}, \btheta) = \sum_{n=1}^{N} W^{(n)} p(\bX_t^{k} \mid \bx_{1:T}^{1:k-1}, \bx_{t+1:T}^{k},\widehat{\bX}_{1:T}^{k+1:K;(n)}, \bY_{1:T}^{k}, \btheta ).
\]
}
\STATE{- Sample $\bx_t^{k}$ from $\widehat{p} (\bX_t^{k} \mid \bx_{1:T}^{1:k-1}, \bx_{t+1:T}^{k}, \bY_{1:T}^{k:K}, \btheta)$ and for $n=1,...,N$ set weight  \[ 
\mbox{ $ W^{(n)} \propto W^{(n)}  p(\bx_t^{k} \mid \bx_{1:T}^{1:k-1}, \bx_{t+1:T}^{k},\widehat{\bX}_{1:T}^{k+1:K;(n)}, \bY_{1:T}^{k}, \btheta ). $ } \]  }
\ENDFOR
\ENDFOR

\end{algorithmic}

}
}
}

\end{algorithm}

\begin{algorithm}[t]

{ \spacingset{1} 
{\footnotesize

\caption{Calculate modified forward filtered probabilities for individual $k$ and sample $n$ in MIFFBS.}
\label{A:MFFP}

{

\begin{algorithmic}
\IF{$t=1$}
\STATE{- Set initial probabilities $P^k_1(\bX_1^{k}) = p(\bX_1^{k} \mid \btheta )$ for $\bX_1^{k} \in \mathcal{S}$.}
\ELSE
\STATE{- Calculate predictive probabilities for $\bX_t^{k} \in \mathcal{S}$
\[ P^k_t(\bX_t^{k}) = \sum_{s=1}^{S} p(\bX_t^{k} \mid \bx_{t-1}^{1:k-1},\bX_{t-1}^{k}=s,\widehat{\bX}_{t-1}^{k+1:K;(n)},\btheta ) p(\bX_{t-1}^{k}=s \mid \bx_{1:t}^{1:k-1},\widehat{\bX}_{1:t}^{k+1:K;(n)}, \bY_{1:t-1}^{k}, \btheta ). \]
}
\ENDIF
\IF{$t<T$}
\STATE{- Calculate the product of the transition probabilities of remaining individuals for $\bX_t^{k} \in \mathcal{S}$
\[ Q_{t}^k(\bX_t^{k}) = p(\bx_{t+1}^{1:k-1},\widehat{\bX}_{t+1}^{k+1:K;(n)} \mid \bx_{t}^{1:k-1}, \bX_t^{k}, \widehat{\bX}_{t}^{k+1:K;(n)}, \btheta  ) \]
}
\ELSE
\STATE{- Set $Q^k_T(\bX_T^{k}) = 1$ for $\bX_T^{k} \in \mathcal{S}$.}
\ENDIF
\STATE{- Calculate modified forward filtered probabilities for $\bX_t^{k} \in \mathcal{S}$
\[ p(\bX_t^{k} \mid \bx_{1:t+1}^{1:k-1},\widehat{\bX}_{1:t+1}^{k+1:K;(n)}, \bY_{1:t}^{k}, \btheta ) = p(\bY_t^k \mid \bX_t^{k}, \btheta )  Q_{t}^k(\bX_t^{k}) P^k_t(\bX_t^{k}). \]
}
\end{algorithmic}

}
}
}

\end{algorithm}

\subsection{Considerations for implementation}

IFFBS can be coded to give scaling $\mathcal{O}(S^3 K T)$ or $\mathcal{O}(S K^2 T)$. 
In this paper we are primarily interested in the scaling properties as the number of chains increases, and so use the former construction.
This requires tracking appropriate summary statistics, which in general would be the number of chains in each state at any time, in order to avoid repeatedly summing over every chain.
For applications with a small number of chains, the second construction may be more efficient.

For some models the suggested proposal distributions will be invalid if they lack full posterior support.
In infectious disease models, for example, DIFFBS will provide a valid proposal distribution if there is an exterior source of infection pressure, but might not otherwise.
In particular, consider DIFFBS being initialized on a realization of the hidden state process in which individual 1 is the only infected individual over some time interval $\mathcal{T}$, after which others are infected. 
Proposed values for individual 1 will always include a period of infection over $\mathcal{T}$, as without this there is zero probability of other individuals becoming infected at later times.
For MIFFBS the probability of constructing an invalid proposal distribution decreases as the number of guiding samples increases.

\section{Application: Avian influenza in chickens}
\label{S:AIV}

Our motivating example is a study of highly pathogenic avian influenza (HPAI) in chickens \citep{Lyall2011}.
Chickens were genetically modified (transgenic chickens) to expresses a short-hairpin RNA that interferes with virus propagation, providing resistance to HPAI.
In order to test the efficacy of the modification, a series of transmission experiments were undertaken.
A crossed experimental design was employed with four independent experiments.
In each experiment there are five \textit{challenge} chickens (inoculated with HPAI at the start of the experiment), and twelve uninfected \textit{in-contact} chickens.
Group 1 consisted of five non-transgenic challenge chickens and twelve non-transgenic in-contact chickens, group 2 consisted of five non-transgenic challenge chickens and twelve transgenic in-contact chickens,  group 3 consisted of five transgenic challenge chickens and twelve non-transgenic in-contact chickens, and group 4 consisted of five transgenic challenge chickens and twelve transgenic in-contact chickens.
Each experiment was undertaken over 10 days, with the chickens being observed twice daily.
Chickens were removed from the experiment for a variety of reasons: dying naturally, removed for being clinically sick, and removed at random for immunohistological studies.
The individual-level data can be ascertained in Figure~\ref{F:StatePost}.

\subsection{Aims}

The \cite{Lyall2011} experiments demonstrate a clear difference between transgenic and non-transgenic chickens, with non-transgenic chickens having a larger mortality rate than transgenic chickens.
However, it is not immediately clear if this difference results from differences in transmission potential, susceptibility to infection, or both.
\cite{McKinley2020} endeavors to resolve this ambiguity by using Bayesian model selection with population-level data, specifically the number of removed chickens of each type per half day.
Visibly sick (moribund) birds are artificially removed from the experiment, but are assumed to have died naturally within the next half day.
The authors use a technique called Approximate Bayesian Computation (ABC), which targets an approximate posterior distribution $\pi_{\epsilon} \left( \btheta \mid \bY_{1:T} \right)$ and evaluates an approximate model evidence $\pi_{\epsilon} \left( \bY_{1:T} \right)$.
Broadly speaking, ABC involves generating a set of simulated data $\bY^{sim}_{1:T}$ from the model whenever a parameter value is proposed, and accepting the proposal whenever the distance between simulated and observed data is smaller than some threshold, $\epsilon$.
By implementing an  ABC version of the alive particle filter (APF) within particle marginal Metropolis Hastings (PMMH) and constructing novel efficiency saving methods, \cite{McKinley2020} achieves a small threshold, where at each observation the cumulative number of observed deaths can differ by at most 1.
The resulting Bayes factors suggest that the transmission rate differs between transgenic and non-transgenic chickens, but that the initial infection probability, death rate, and susceptibility are the same.

Our goal here is to improve on the analysis of \cite{McKinley2020} by using individual-level models and targeting the correct posterior  distribution. 
By using individual-level models, we are able to make full use of the available data rather than condensing the data into population-level summaries.
This may be more informative for our inference.
Likewise, we may obtain more reliable conclusions by targeting the correct posterior distribution rather than an approximation.

\subsection{Models}

The models under consideration in \citet{McKinley2020} are continuous-time SIR models, where at any point in time each chicken may be susceptible (S), infected (I), or removed (R). 
The simplest model treats transgenic and non-transgenic chickens identically.
Challenge birds are infected on day 0 with probability $p$.
The probability of an $S \rightarrow I$ transition over some small time interval $[t, t+dt)$ is
\[ P (S \rightarrow I \mbox{ in } [t, t+dt)) = \frac{\beta S I}{N} dt + o(dt), \]
\noindent where $S$ denotes the number of susceptible chickens, $I$ denotes the number of infected chickens, $N$ denotes the total number of chickens, and $\beta$ is the transmission parameter. Likewise the probability of an $I \rightarrow R$ transition over some small time interval $[t, t+dt)$ is
\[ P (I \rightarrow R \mbox{ in } [t, t+dt)) = \gamma I dt + o(dt), \]
\noindent where $\gamma$ is the removal rate.

The alternative models under consideration contain different parameters for transgenic and non-transgenic chickens.
For instance, the probability of infection on day 0 for challenge birds may be $p^{N}$ for non-transgenic chickens and $p^{T}$ for transgenic chickens.
Likewise there may be two transmission parameters ($\beta^{N}$ and $\beta^{T}$), two removal rates ($\gamma^{N}$ and $\gamma^{T}$),  an additional susceptibility term ($\nu^{N}$),
or any combination of these.
The parameterizations for all 16 models are shown in Table \ref{T:ModelList}.

In order to apply our methods we need discrete-time individual-level model equivalents to the continuous-time population-level models presented in \citet{McKinley2020}.
To do this, we start with the continuous-time model and make the assumption that for each half-day period the transition rates remain constant.
In other words, if a chicken becomes infected within a half day period, it does not become infectious until the start of the next half-day period.
We can then directly evaluate the transition probability matrix for each chicken over each half day.
For example, in the case of the simplest SIR model the transition probability matrix for any chicken from time $t$ to $t+0.5$ is
\begin{equation*}
\mathbb{P}_t = \left( \begin{array}{c c c} E_{SI} & R_g \left( E_{IR} - E_{SI} \right) & 1 + \left( R_g - 1 \right) E_{SI} - R_g E_{IR} \\
 0 & E_{IR} & 1 - E_{IR} \\ 0 & 0 & 1  \end{array} \right),
\end{equation*}
\noindent where
\begin{equation*}
E_{SI} = \exp \left( - 0.5 \frac{\beta I_t}{N} \right), \; \; \; E_{IR} = \exp \left( - 0.5 \gamma \right), \; \; \; R_g = \frac{\frac{\beta I}{N}}{\frac{\beta I}{N} - \gamma}, \\
\end{equation*}
\noindent and $I_t$ is the number of infected chickens at time $t$.

Following \cite{McKinley2020}, our prior distributions are $\mathcal{U}(0,1)$ for $p^N$ and $p^T$, and $\mbox{Exp}(1)$ for $\beta^N$, $\beta^T$, $\nu^N$, $\gamma^N$, and $\gamma^T$. 
Since our primary goal in this section is to compare Bayes factor estimates from our analyses with those presented in \cite{McKinley2020}, 
and that our focus is on the methodology for constructing proposal distributions for the hidden state process,
we do not consider the issue of sensitivity to the prior distributions here.
However, we do stress that Bayes factors are sensitive to the choice of prior distributions, and model selection analyses should be mindful of this.

\begin{table}[t]
\centering
\begin{tabular}{c c c c c c c c c c c c c c c c c} 
\cline{1-5}
\cline{7-11}
\cline{13-17}
Model & \multicolumn{4}{c}{Parameters} & & Model & \multicolumn{4}{c}{Parameters} & & Model & \multicolumn{4}{c}{Parameters} \\
 & $p$ & $\beta$ & $\nu$ & $\gamma$ &  & & $p$ & $\beta$ & $\nu$ & $\gamma$ &  &  & $p$ & $\beta$ & $\nu$ & $\gamma$ \\
\cline{1-5}
\cline{7-11}
\cline{13-17}
 Model 1: & 1 & 1 & 0 & 1  & & Model 7: & 1 & 2 & 1 & 1 & & Model 13: & 1 & 1 & 1 & 2 \\
 Model 2: & 2 & 1 & 0 & 1  & & Model 8: & 2 & 2 & 1 & 1 & & Model 14: & 2 & 1 & 1 & 2 \\ 
 Model 3: & 1 & 2 & 0 & 1  & & Model 9: & 1 & 1 & 0 & 2 & & Model 15: & 1 & 2 & 1 & 2 \\ 
 Model 4: & 2 & 2 & 0 & 1  & & Model 10: & 2 & 1 & 0 & 2 & & Model 16: & 2 & 2 & 1 & 2 \\ 
 Model 5: & 1 & 1 & 1 & 1  & & Model 11: & 1 & 2 & 0 & 2 & & & & & &  \\  
 Model 6: & 2 & 1 & 1 & 1  & & Model 12: & 2 & 2 & 0 & 2 & & & & & &  \\ 
\cline{1-5}
\cline{7-11}
\cline{13-17}
\end{tabular}
\caption{
\label{T:ModelList}The number of parameters in each of the 16 models. For the initial infection ($p$), transmission ($\beta$), and removal parameters ($\gamma$), a single parameter indicates that there is no distinction between transgenic and non-transgenic birds and two parameters indicates a distinction. Likewise for the susceptibility parameter ($\nu$), zero parameters indicates no distinction and one parameter indicates a distinction.}
\end{table}

\subsection{Results}
\label{S:Results}

\subsubsection{Simulated data}

We compare the bias, precision, and scaling of DIFFBS, MIFFBS, Chib's method \citep{Chib1995} and a particle filter \citep{Doucet2011} on simulated data.
Since the primary challenge in obtaining model evidence estimates lies in integrating over the hidden state process,
we restrict this part of the investigation to determining the marginal likelihood conditioned on a fixed set of parameter values.

We generate 5 sets of simulated data from model 16, each using the same cross experimental design as the real data, but with varying numbers of chickens.
Specifically, we simulate data with 4, 8, 16, 32, and 64 chickens per pen, with 1, 2, 5, 10, and 19 challenge birds respectively.
The parameters for simulating the data are $p^N = 0.9$, $p^T = 0.8$, $\nu^N = 1.2$, $\beta^N = 2.3$, $\beta^T = 1.4$, $\gamma^N = 0.5$, and $\gamma^T = 0.3$,
which are then fixed for the purpose of estimating the marginal likelihood.
We censor some chickens in the 12 hours preceding a death, to mimic removals from finding a moribund bird.
In particular, whenever we have simulated a death, the chicken is removed early with probability 0.5.
We tune each of the methods to generate 1000 estimates of the marginal likelihood in one hour for each data set. 

For DIFFBS we first obtain 100 guiding samples, from which we determine a high posterior realization $\widetilde{\bX}_{1:T}$ on which to condition our proposal.
We repeatedly propose from this same realization for a set number of replicates, which acts as our tuning variable to control the computational cost.
The average of the replicated estimates then provides one estimate of the marginal likelihood.

For MIFFBS the tuning variable is the number of guiding samples used to construct the proposal distribution.
The guiding samples are regenerated whenever the effective sample size has fallen to half the initial sample size.
For convenience, we only perform this regeneration whenever we start proposing values for a new individual.

In order to apply Chib's method, we need to define parameter blocks for which we have the full conditional distributions.
We define each block as an individual update, as we do in DIFFBS and MIFFBS.
Chib's method approximates the full conditional distribution for the hidden state process via a series of mixture distributions.
Similar to MIFFBS, Chib's method is guided by samples of the hidden state process,
the number of which again acts as our tuning variable.
In \cite{Chib1995} any realization of the hidden state process can be used as an input into the approximation for a marginal likelihood estimate,
but here we sample from the approximation as an importance proposal.
Chib's method shares the same possible bias issues as DIFFBS and MIFFBS, but as with MIFFBS the probability of this occurring decreases as the number of guiding samples increases.

The particle filter operates sequentially through time, and requires the specification of proposal distributions that are ideally conditioned on upcoming observations.
Simply using observations of whether chickens are dead or alive leads to particle degeneracy, especially with larger numbers of chickens.
The reason for this is the removal of moribund birds.
Moribund birds are observed as alive upon being extracted from the pen, and so may be sampled as susceptible.
However, an extracted chicken can not then transition to the observed $R$ state as it can no longer become infected.
It is clear that moribund birds must be infected at these extraction times, and we must condition our proposal distributions accordingly.
The particle filter provides unbiased estimates of the marginal likelihood, and the tuning variable is the number of particles.

Finally, when we have a small number of chickens (4 or 8 chickens per pen), we can compare our estimates with the true marginal likelihood, which can be obtained using the forward filter.
This takes 0.04 seconds with 4 chickens per pen, and 262 seconds with 8 chickens per pen.
The computational cost scales by a factor of 9 per additional chicken, and so the cost with 16 chickens per pen would be 358 years, demonstrating the poor scalability of the forward filter when applied to individual-level models.

\begin{table}[p]
\centering
\begin{tabular}{c c c c c c c c} 
\hline
Chickens & &  & \multicolumn{5}{c}{Log marginal likelihood} \\ [6pt] 
Per Pen &  &  & FF & DIFFBS & Chib's & MIFFBS & PF \\
   \hline
    4 & Mean & $-25.$ & 7860 & 8042 & 7865 & 7866 & 7860 \\
    & Range & $-25.$ & & $[8094,7990]$ & $[7883,7846]$ & $[7875,7857]$ & $[7867,7854]$ \\
       \hline
   8 & Mean & $-69.$ & 4389 & 4712 & 4401 & 4379 & 4390 \\
    & Range & $-69.$ &  & $[4813,4612]$ & $[4468,4334]$ & $[4426,4369]$ & $[4404,4377]$ \\
           \hline
   16 & Mean & $-175.$ &  & 9546 & 8970 & 8959 & 8958   \\
    & Range & $-175.$ &  & $[9661,9432]$ & $[9104,8837]$ & $[9004,8914]$ & $[9011,8905]$ \\
           \hline
   32 & Mean  & $-350.$ &   & 1423 & 1288 & 1347 & 1323  \\
     & Range & $-350.$ & & $[1487,1359]$ & $[1444,1135]$ & $[1395,1300]$ & $[1427,1220]$ \\
            \hline
   64 & Mean & $-706.$ &  & 8932 & 8788 & 8897 & 8911   \\
    & Range & $-706.$ &  & $[9010,8854]$ & $[9059,8525]$ & $[8957,8838]$ & $[9418,8429]$ \\
    \hline
\end{tabular}
\caption{Log marginal likelihood estimates in the simulation study for DIFFBS, Chib's method, MIFFBS, and the particle filter. For 4 and 8 chickens per pen we have also included the true log marginal likelihood obtained from the forward filter. The integer component of the log marginal likelihood is shown to the left of the methods. The mean estimate is the log of the average marginal likelihood from 1000 estimates. The range adds/subtracts 3 standard errors.}
\label{T:SS_Comp_Full}
\end{table}

The results are presented in Table \ref{T:SS_Comp_Full}, which includes the mean marginal likelihood estimates and ranges indicated by $\pm 3$ standard errors.
Comparing the mean estimates to the true values, DIFFBS clearly has a small bias.
With both 4 and 8 chickens per pen, the true values does not lie within 3 standard errors of the mean.
The remaining methods seem to perform well, with the mean estimates being close to the true value.
Even though Chib's method and MIFFBS have the potential to be biased, there seem to enough guiding samples to prevent this issue.

Comparing the different methods, the particle filter provides the most precise estimates in the smaller data sets (4 and 8 chickens per pen), and MIFFBS provides the most precise estimates in the larger data sets (16, 32, and 64 chickens per pen).
This is most notable with 64 chickens per pen, where the range of the particle filter estimates is $\sim 0.1$ on the log scale, vs $\sim 0.01$ for MIFFBS. 
Chib's method is more precise than the particle filter in the largest dataset, but less precise than both MIFFBS and the particle filter otherwise.
Finally, DIFFBS maintains a relatively good precision, but as stated provides biased estimates.

Broadly speaking, DIFFBS and the particle filter scale linearly in the number of individuals, meaning that the tuning variables need to be halved for a doubling of the population size in order to maintain the same computational cost.
Some factors do impact this, including computations in the algorithms that are not impacted by the tuning variable, and run times also have a stochastic element.
DIFFBS uses 8346 replicates in the smallest data set and 645 in the largest, and the particle filter uses 189 405 particles in the smallest data set and 11 900 in the largest.
Chib's method scales quadratically in the number of individuals, and so in the largest data set has only 18 samples constructing the proposal distribution for any individual, down from 1096 in the smallest.
The scaling of MIFFBS depends on the frequency of the guiding sample regeneration.
If the number of regeneration steps is proportional to the number of individuals, then MIFFBS will scale quadratically in the number of individuals.
If, on the other hand, no regeneration is required, MIFFBS scales linearly in the number of individuals.
With adaptive regeneration scheme it is also possible for the scaling to land somewhere between these two extremes.
MIFFBS used 4695 guiding samples in the smallest data set, and 428 in the largest.
This is similar to the scaling observed in DIFFBS, suggesting near-linear scaling for this investigation.

\subsubsection{HPAI data}

We apply MIFFBS within importance sampling to estimate the model evidence for all 16 models fitted to the individual-level HPAI data.
For the proposal distribution on the model parameters we construct a defense mixture, which was shown to perform well in \cite{Touloupou2018}.
For each model we run an MCMC algorithm targeting the joint posterior distribution $\pi (\btheta, \bX_{1:T} \mid \bY_{1:T})$.
We obtain 10000 samples in each case, which takes approximately 30 seconds and does not significantly contribute to the cost of the model selection algorithm. We then transform the model parameters as $\log(-\log(p^N))$, $\log(-\log(p^T))$, $\log(\nu^N)$, $\log(\beta^N)$, $\log(\beta^T)$, $\log(\gamma^N)$, and $\log(\gamma^T)$, to which we fit a multivariate Gaussian distribution.
The defense mixture is then a weighted mixture distribution of the prior and the fitted Gaussian, with weights 0.05 and 0.95 respectively.

Each time a new set of parameters are proposed we generate 1000 guiding samples of the hidden state process using IFFBS.
We use an adaptive regeneration scheme for the guiding samples, which is triggered whenever the effective samples size is smaller than 500 as we initiate a proposal for a new individual.

\begin{table}[p]
\centering
\begin{tabular}{c c c l c c r l} 
\hline
  & & \multicolumn{3}{c}{MIFFBS / Individual level} & \multicolumn{3}{c}{ABC / Population level}  \\
  & & Mean & & Range & &  \\
  \hline
  Model 1 & & $-136.86$ & \color{Green} * & $[-136.94,-136.80]$ &  & $-36.93$ & \color{Green} *  \\ 
  Model 2 & & $-136.94$ & \color{Green} * & $[-137.03,-136.85]$ &  & $-37.27$ & \color{Green} *  \\
  Model 3 & & $-134.64$ & \color{red} *** & $[-134.88,-134.45]$ &  & $-35.64$ & \color{red} ***  \\
  Model 4 & & $-134.80$ & \color{blue} ** & $[-135.06,-134.59]$ &  & $-35.93$ & \color{blue} ** \\
  \hline
  Model 5 & & $-137.94$ &  & $[-138.01,-137.87]$ &  & $-37.71$ & \color{Green} * \\ 
  Model 6 & & $-137.85$ &  & $[-137.97,-137.75]$ &  & $-38.06$ &  \\ 
  Model 7 & & $-135.69$ & \color{blue} ** & $[-135.89,-135.52]$ &  & $-36.51$ & \color{blue} ** \\
  Model 8 & & $-135.67$ & \color{blue} ** & $[-135.89,-135.50]$ & &  $-36.71$ & \color{blue} **  \\
  \hline
  Model 9 & & $-138.09$ &  & $[-138.19,-138.00]$ &  & $-37.85$ & \color{Green} *  \\
  Model 10 & & $-138.09$ &  & $[-138.26,-137.94]$ &  & $-38.17$ &   \\ 
  Model 11 & & $-135.52$ & \color{blue} ** & $[-135.72,-135.35]$ &  & $-36.61$ & \color{blue} ** \\ 
  Model 12 & & $-135.37$ & \color{blue} ** & $[-135.63,-135.16]$ &  & $-36.86$ & \color{Green} * \\ 
  \hline
  Model 13 & & $-139.03$ &  & $[-139.24,-138.85]$ & &  $-38.44$ &   \\ 
  Model 14 & & $-139.10$ &  &$[-139.28,-138.95]$  &  & - &   \\ 
  Model 15 & & $-136.47$ & \color{Green} * &$[-136.71,-136.28]$  & &  $-37.41$ & \color{Green} *  \\ 
  Model 16 & & $-136.35$ & \color{Green} * & $[-136.60,-136.15]$ & &  $-37.74$ & \color{Green} *  \\ 
  \hline
\end{tabular}
\caption{Comparison of model evidence estimates and rankings for the MIFFBS importance sampling algorithm on the individual level model and the ABC alive particle filter on the population level model \citep{McKinley2020}. The mean column gives the log of the model evidence estimate, and range indicates the log model evidence with $\pm 3$ standard errors. {\color{red} ***} Indicates the best model, {\color{blue} **} indicates models with Bayes between 1 and $1 / 3.2$ with respect to the best model, and {\color{Green} *} indicates models with Bayes between $1 / 3.2$ and $1 / 10$.}
\label{T:Full}
\end{table}

The model evidence estimates are shown in Table \ref{T:Full} compared with those provided in \citet{McKinley2020}.
The model evidence estimates do not directly compare, as \citet{McKinley2020} uses population-level models and data, whereas we use individual-level models and data.
However, we can compare the rankings, and the strength of evidence between models indicated by the Bayes factors.
Model 3 is favored in both analyses, which indicates two transmission parameters, one initial infection parameter, one removal parameter, and zero susceptibility parameters.
Considering a threshold for the Bayes factor of 3.2 for substantial evidence in favor of the superior model \citep{Kass1995}, both analyses find models 4, 7, 8, and 11 as having significant support in comparison to model 3, although the exact ordering changes slightly.
We also find significant support for model 12.
With a threshold of 10 to indicate strong evidence in favor of the superior model, our analysis is more discriminating. 
We find strong evidence against six models, whereas \citet{McKinley2020} finds strong evidence against four.
The ABC algorithm failed to estimate a model evidence for model 14, due to requiring an excessively large number of model simulations.
We have no issues with this, but do determine it to be the weakest model.

%The similarities between the results are reinforced by comparing the model averaged posterior weights for the number of parameter of each type.
%We find a posterior probability of 0.91 for two transmission parameters vs 0.79 in \citet{McKinley2020}, a probability of 0.72 for zero susceptibility parameters vs 0.7, a probability of 0.68 for one removal parameter vs 0.72, and  a probability of 0.51 for one initial infection parameter vs 0.58.

\begin{figure}[t]
\centering
\includegraphics[width=1\textwidth]{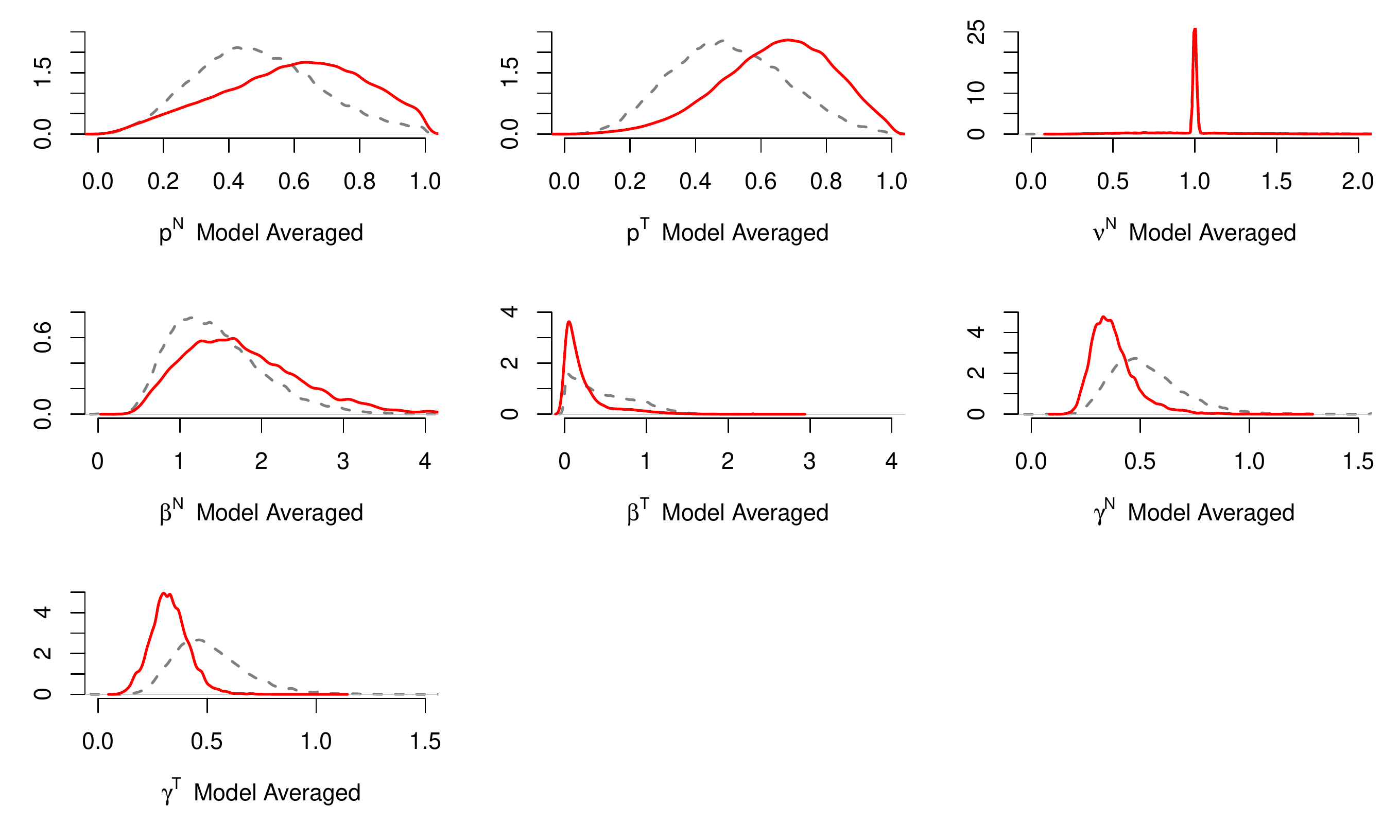} 
\caption{Model averaged posterior distributions for the model parameters. Dashed lines are those presented in \citet{McKinley2020}, and solid lines are those obtained from the individual-level analysis using MIFFBS.}
\label{F:ParmPost}
\end{figure}

The model averaged posterior distributions are shown in Figure~\ref{F:ParmPost}, compared with those in \citet{McKinley2020}. 
There is significant overlap in the marginal posterior distribution for each parameter, but we estimate larger values for the initial infection probabilities $p^N$ and $p^T$, and smaller values for the removal parameters $\gamma^N$ and $\gamma^T$.

\begin{figure}[p]
\centering
\includegraphics[width=1\textwidth]{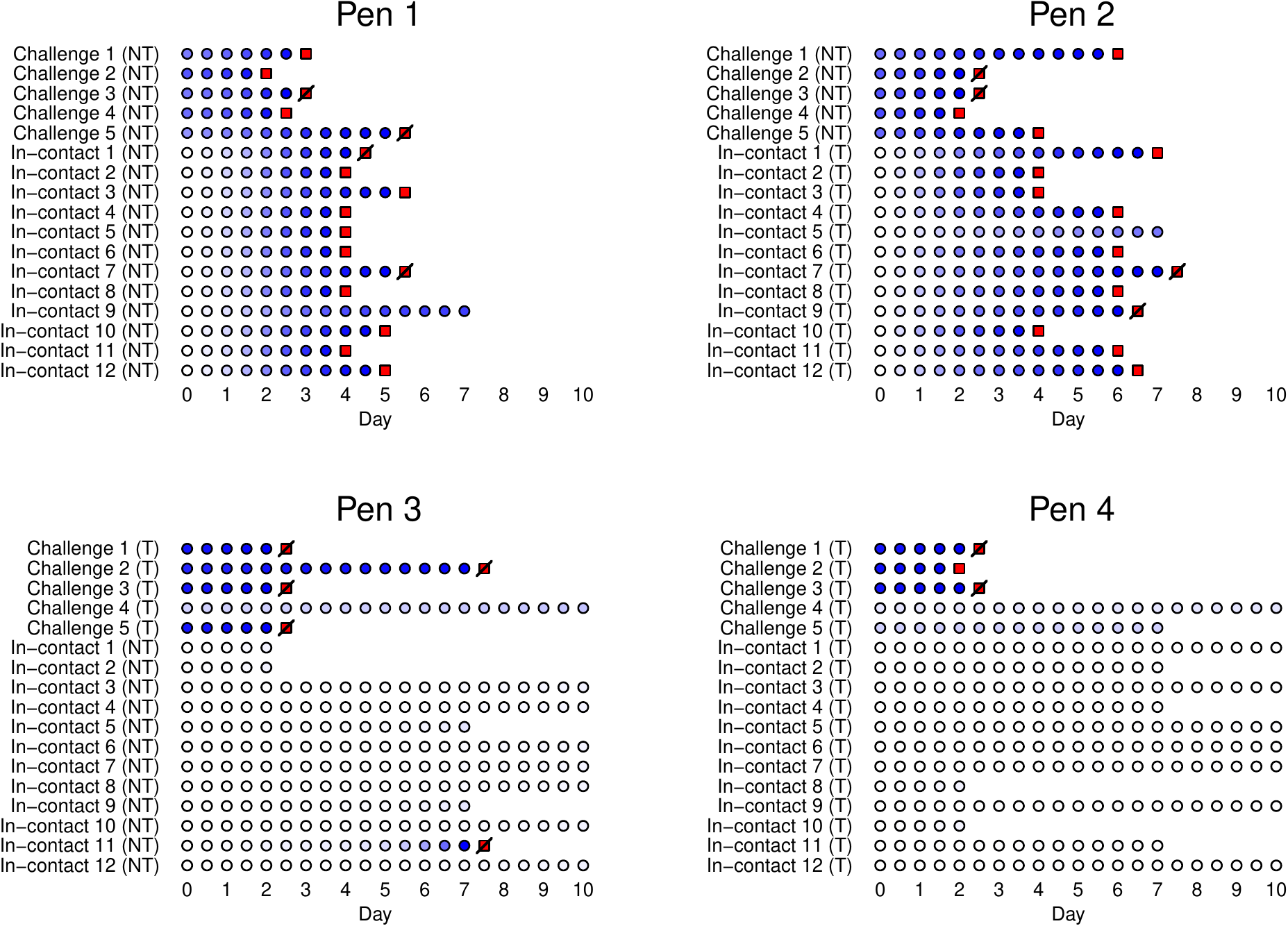} 
\caption{Marginal posterior daily infection probabilities for each chicken. Shaded circles indicate the infection probabilities (white indicates probability zero). 
Squares indicates a known death either via direct observation (no strike) or implied via removal of moribund birds (with strike). }
\label{F:StatePost}
\end{figure}

An advantage of using individual-level models and data is that we can infer information about each individual through time.
The model averaged marginal state posterior distributions are shown in Figure \ref{F:StatePost}, along with the observed removal times.

%Both the state and parameter posteriors change strongly depending on the model, and are shown for each model in the supplementary information {\color{red} REFERENCES}.

\section{Discussion}
\label{S:Discussion}

In this paper we have introduced effective algorithms for proposing values for the hidden state process of CHMMs with weak inter-chain dependencies, which we refer to as DIFFBS and MIFFBS.
These proposal distributions are based on a chain-level decomposition, which naturally takes advantage of the dependency structure, providing algorithms that remain computationally viable as the number of chains increases and that require little user input.

We demonstrated the utility of these methods in both simulation studies and by undertaking a model selection experiment for highly pathogenic avian influenza.
This data set has previously been investigated in \citet{McKinley2020} at the population level with an approximate inference scheme.
We were able to improve on this study by better utilizing the data with individual-level models, and by targeting the correct posterior distribution. 
The key biological finding from \citet{McKinley2020} is strong evidence for a difference in transmission between transgenic potential and non-transgenic chickens, and our results give refined, but consistent conclusions.

There are several possible extensions to the proposed methodology.
DIFFBS was shown to be biased in our application, and MIFFBS can potentially be biased if a small number of guiding samples are used.
There are potential solutions to explore for this issue.
For example, the transition probabilities in the proposal distributions could be modified to ensure full posterior support, or else DIFFBS and MIFFBS could be included as part of a mixture distribution.
Since any such modification will lead to a worse approximation of the full conditional distribution, these choices are likely to lead to bias-variance trade-offs.

DIFFBS and MIFFBS will lose efficiency when the chains are strongly correlated.
If correlated groups of chains can be determined, for instance by picking out households in an infectious disease model, then block updates may be used in place of individual-level updates.
Here a modified forward filter can be used to estimate the full conditional distribution of the hidden state process for the block of chains, conditional on the remainder.
However, the computational cost scales exponentially with the size of the largest block.

Focusing on MIFFBS, here we have only considered regenerating the guiding samples upon initializing a proposal for a new chain.
This simplifies the implementation of the algorithm, but could be further generalized.
This will be important in problems with long time horizons, where we can expect degeneracy to occur in the weights.

Whilst we have focused on model selection, MIFFBS could also be useful in estimating model parameters.
For example, in infectious disease models, parameter values and the hidden state process are frequently strongly correlated, and so MCMC implementations that iterate between updating the model parameters and hidden states do not mix effectively. 
MIFFBS could overcome such difficulties if used within a pseudo-marginal scheme \citep{Andrieu2009} to update both components jointly.

In conclusion, we have presented novel methods for sampling the hidden state process of CHMMs with weak inter-chain dependencies, with favorable scaling properties as the number of chains increases.
These sampling distributions are constructed automatically from the model specification and data, requiring little user input or tuning.
This is a particularly useful feature when performing model selection, as the algorithms do not need to be modified to target different models.
These properties have been illustrated in a model selection investigation of highly pathogenic avian influenza data, and the methods are readily implementable to many other infectious disease models and wider-ranging applications.

%\bigskip
%\begin{center}
%{\large\bf SUPPLEMENTARY MATERIAL}
%\end{center}

%The supplementary materials provide code for reproducing the results. 

\bibliographystyle{agsmdoi}

\bibliography{MiFFBSBib}
\end{document}